\documentclass[preprint,prd,noshowpacs,nofootinbib]{revtex4}

\usepackage{color} 
\usepackage{amssymb}
\usepackage{amsmath}
\usepackage{graphicx} 
\usepackage{float}
\usepackage{braket}     
\usepackage{pdfpages}
\usepackage{epstopdf}
\usepackage[utf8]{inputenc}

\begin{document}

\title{Temporal Pound-Rebka experiment as gravitational Aharonov-Bohm effect \footnote{Essay written for the Gravity Research Foundation 2024 Awards for Essays on Gravitation. \\
~ \\}}

\author{RY Chiao}
\email{raymond_chiao@yahoo.com}
\affiliation{University of California, Merced,  School of Natural Sciences,
Merced, CA 95344, USA}
\author{NA Inan}
\email{ninan@ucmerced.edu}
\affiliation{Clovis Community College, 10309 N. Willow, Fresno, CA 93730 USA}
\affiliation{University of California, Merced, School of Natural Sciences, 
Merced, CA 95344, USA}
\affiliation{Department of Physics, California State University Fresno, Fresno, CA 93740-8031, USA}
\author{DA Singleton \footnote{Corresponding Author}}
\email{dougs@mail.fresnostate.edu}
\affiliation{Department of Physics, California State University Fresno, Fresno, CA 93740-8031, USA}
\author{ME Tobar}
\email{michael.tobar@uwa.edu.au}
\affiliation{Quantum Technologies and Dark Matter Labs, Department of Physics, University of Western Australia, Crawley, WA 6009, Australia.}

\date{\today}

\begin{abstract}
One of the classical tests of general relativity is the precision measurements by Pound and Rebka of red-shift/blue-shift of photons in a gravitational field. In this essay, we lay out a temporal version of the Pound-Rebka experiment. The emission and absorption of photons occurs at different times, rather than at different spatial locations as in the original Pound-Rebka experiment. This temporal Pound-Rebka experiment is equivalent to a gravitational Aharonov-Bohm Effect and is testable via current or near future satellite experiments.
\end{abstract}

\maketitle

{\bf Pound-Rebka and Aharonov-Bohm effect:}
The Pound-Rebka experiment \cite{pound} is one of the classic, experimental tests of general relativity, along with the deflection of starlight and the perihelion precession of the orbit of Mercury. Pound and Rebka sent $\gamma$-ray photons between an emitter and a receiver separated by about 20 meters vertically in the Earth's gravitational field. The $\gamma$-rays going from bottom to top were red-shifted while those going from top to bottom were blue-shifted. The red and blue shifts over this distance were extremely small, but Pound and Rebka used the M{\"o}ssbauer effect \cite{mossbauer}, which allowed for amazingly precise determination of small frequency shifts due to the recoil-less emission and absorption of the $\gamma$-rays. Using the M{\"o}ssbauer effect Pound and Rebka we able to distinguish even the small red/blue shift that occurs over just tens of meters.  

In this essay, we propose a temporal version of the Pound-Rebka experiment where the emitter and receiver are at almost the same spatial location inside a satellite, but the gravitational potential is varied with time. This is achieved by placing the Pound-Rebka setup in a satellite in a low Earth, slightly elliptical orbit. In this way, the M{\"o}ssbauer apparatus will experience different gravitational potentials at different points of the orbit.    

Here we claim that this temporal Pound-Rebka experiment is equivalent to the gravitational Aharonov-Bohm (AB) effect \cite{AB} -- specifically that it is the gravitational analog of the scalar-electric AB effect. Unlike the more well-known vector-magnetic AB effect, the scalar-electric AB effect involves splitting up a beam of charged, quantum particles and sending them through Faraday cages. While the particles are inside the Faraday cages \footnote{In the proposal of reference \cite{AB} the Faraday cages were long metal tubes.} a time-varying potential difference was applied to the tubes. This resulted in a measurable phase shift when the particles were recombined after exiting the Faraday tubes. 

Reference \cite{chiao-2023} proposed a new approach to the scalar-electric AB effect which avoided the difficulty in timing the turn-on/turn-off of the potential difference with the entry/exit of the quantum particles from the Faraday tubes. In this new version of the scalar-electric AB effect a quantum system \footnote{In reference \cite{chiao-2023} a hydrogen-like atom like rubidium was proposed.} is placed in a single Faraday shell which had a sinusoidal varying potential. The quantum system inside the Faraday shell would see the time-varying potential, but no electric field. Despite this there would be a shift in the energy level spectrum of the quantum system. This is in contrast to the standard scalar-electric AB effect where it is a shift in the interference pattern that signals the AB effect. We now show that a similar setup is possible to probe the gravitational AB effect.

{\bf Gravitational Aharonov-Bohm Effect:} 
The scalar-electric AB effect creates a changing electric potential, $\Phi _e (t)$, by moving charges on to and off of the Faraday shell. This is not feasible for the gravitational interaction since the gravitational interaction is much weaker than the electromagnetic interaction, and the amount of mass that could be moved onto and off of the shell is small. Thus for this setup, the gravitational AB effect would be unobservable with current technology. To observe the gravitational AB effect we obtain a varying gravitational potential by varying the distance between the quantum system and the source of the gravitational field. This gives the gravitational potential of the form    
\begin{equation}
\Phi _g (  t )     =-\frac{G M }{r (t)}~.
\label{Ab-grav-2}
\end{equation}
$G$ is Newton's constant, $M$ is the mass of the source of the gravitational field about which the quantum system orbits, and  $r (t)$ is the time-dependent distance between the quantum system and the mass $M$. 

The gravitational potential in \eqref{Ab-grav-2} is obtained by placing our quantum system in a satellite in a low-Earth, almost circular orbit. For such an orbit, $r(t)$, can be approximated as a simple oscillatory term plus a constant. The parameters of such orbits are: (i) The mass of the Earth, $M=5.97 \times 10^{24}$ kg. (ii) The perigee and apogee radius from the center of the Earth which are $r_p = 6.800 \times 10^6$ m and $r_a = 6.810 \times 10^6$ m, respectively. This corresponds to a perigee altitude of 400 km and apogee altitude of 410 km given that the Earth's radius is $r_E \approx 6400$ km. These radii correspond roughly to those of the International Space Station (ISS). (iii) The period of a satellite with this apogee/perigee is about $T \approx$ 90 minutes or 5400 seconds giving an angular frequency of $\Omega = \frac{2 \pi}{T}  = 1.0 \times 10^{-3} \rm{\frac{rad}{sec}}$ (or $f= \frac{1}{T} = 1.85 \times 10^{-4} ~ {\rm Hz}$). The radius of the orbit as a function of time can be approximated as \footnote{This treatment of nearly circular orbits is essentially that found in section 9.5 of reference \cite{kk}.}
\begin{equation}
    \label{radius}
    r(t) = \frac{r_p+r_a}{2} + \frac{r_p-r_a}{2} \cos(\Omega t) \equiv A + B \cos (\Omega t)~.
\end{equation}
Using the $r_p$ and $r_a$ from above, the $A$ and $B$ parameters defined in \eqref{radius} are $A=6.805 \times 10^6$ m and $B = -5.000 \times 10^3$ m. Perigee occurs at $t=0$ and apogee at $t=\pi/\Omega$. For the chosen $r_p$ and $r_a$, $A \gg B$ so one can approximate $\frac{1}{r(t)} = \frac{1}{A+B\cos (\Omega t)} \approx \frac{1}{A} \left(1 - \frac{B}{A} \cos (\Omega t ) \right)$. With this the gravitational potential in \eqref{Ab-grav-2} becomes
\begin{equation}
    \label{Ab-grav-4}
    \Phi _g (t) \approx  - \frac{GM}{A} \left[ 1 - \frac{B}{A} \cos (\Omega t ) \right]  ~,
\end{equation}

The AB phase for the scalar AB effect is generically given by the charge of the interaction divided by $\hbar$ times the time integral of the potential of the interaction. For the gravitational interaction the ``charge" is the mass, $m$, of the quantum system placed in the gravitational potential, $\Phi _g$. This gives the gravitational AB phase as 
\begin{equation}
    \label{AB-grav-a}
    \varphi _g (t) = \frac{m}{\hbar} \int _0 ^t  \Phi _g (t') dt' \approx
   - \frac{m}{\hbar} \int _0 ^t  \frac{GM}{A} \left[ 1 - \frac{B}{A} \cos (\Omega t' ) \right] dt' =  -\frac{GmM}{\hbar A} t + \alpha \sin (\Omega t) ~.
\end{equation} 
In the last expression in \eqref{AB-grav-a} we have defined the dimensionless frequency modulation (FM)  parameter $\alpha \equiv \frac{GmMB}{\hbar \Omega A^2}$.
It is the second sinusoidal term in \eqref{AB-grav-a} ({\it i.e.} $\alpha \sin (\Omega t)$) which leads to the gravitational AB phase. The linear in time term ({\it i.e.} $-\frac{GmM}{\hbar A} t$) is a constant gravitational shift in the energy of the quantum system.

Now we carry out an analysis of the quantum system in the presence of this time-varying gravitational potential. Since the quantum system is in free fall, it is effectively screened from the gravitational field and forces, as required for AB effect experiments. We begin by assuming that the solution to our quantum system is known in the absence of the gravitational potential $\Phi _g (t)$. That is we can solve  the time-independent Schr{\"o}dinger $H_0 \Psi _i ({\bf x}) = E_i \Psi_i ({\bf x})$) where $H_0$, $\Psi_i$, and $E_i$ are the Hamiltonian, wave function, and energy eigenvalues, respectively, of the quantum system without $\Phi _g (t)$. Next, we place the quantum system in the gravitational potential \eqref{Ab-grav-4}.  This gives a new Hamiltonian $H = H_0 + m \Phi _g (t)$, for which the time-dependent Schr{\"o}dinger equation for $H$ is
\begin{equation}
\label{tdse}
i\hbar \frac{\partial \psi }{\partial t}=H\psi =\left( H_{0}+m \Phi_g (t) \right) \psi.
\end{equation}
We solve \eqref{tdse} by applying the standard separation-of-variables ansatz of the form
$\psi (\mathbf{x},t)=X(\mathbf{x})T(t)$. In \cite{chiao-2024} it was found that the spatial part of the separation of variables ansatz was just the wave function of $H_0$ {\it i.e.} $X({\bf x})=\Psi _{i }({\bf x})$. The time part was found to be
\begin{eqnarray}
\label{T(t) solution}
T(t)  =\exp \left( -\frac{i}{\hbar }\left(E_{i } - \frac{GmM}{A}\right) t-i\alpha \sin \Omega t\right)
\equiv \exp \left( -\frac{i}{\hbar } {\tilde E}_i t-i\varphi ' _g (t)\right)  ~,  
\end{eqnarray}
where ${\tilde E}_i \equiv E_i - \frac{GmM}{A}$ is the initial energy of the quantum system, $E_i$, plus a constant shift, $\frac{GmM}{A}$, due to the gravitational potential, and $\varphi ' _g (t) \equiv \alpha \sin (\Omega t)$ is the gravitational AB phase, which involves the time-dependent part of the gravitational potential from \eqref{Ab-grav-4}. 
Combining the results for $X({\bf x})$ and $T(t)$ the wave function for $H=H_0 + m \Phi_g (t)$ is 
\begin{equation}
\psi _{i }(\mathbf{x},t)=\Psi _{i }(\mathbf{x})\exp \left( -\frac{i}{\hbar}
{\tilde E}_i t - i\varphi ' _g (t)\right) ~. 
\label{psi}
\end{equation}
This new wave function, $\psi _i$, is the original wave function, $\Psi _i $, with an added AB phase factor, $\exp \left( -i\varphi ' _g (t)\right)$. The original energy, $E_i$, is shifted to ${\tilde E}_i$. The exponential of the AB phase term can be evaluated using the Jacobi-Anger expansions (see details in \cite{chiao-2023,chiao-2024}) giving
$\exp \left( - i\varphi ' _g (t)\right) =\exp \left( - i\alpha \sin \Omega t\right) 
=\sum_{n=-\infty }^{\infty } (-1)^n J_{n}(\alpha )\exp \left( in\Omega t\right)$. Inserting this into \eqref{psi} the wave function now reads
\begin{equation}
\psi _i(\mathbf{x},t) =\Psi _i(\mathbf{x}) \sum_{n=-\infty }^{\infty } (-1)^n J_{n}(\alpha )\exp\left( -\frac{i}{\hbar} ( {\tilde E}_i -n\hbar \Omega ) t \right)
\label{psi-2}
\end{equation}
From equation \eqref{psi-2} one can see that the wave function for \eqref{tdse} is an infinite series of oscillating terms going in integer steps from $n=-\infty$ to $n = +\infty$ with energies given by $E_i^{(n)} = {\tilde E}_i \pm n\hbar \Omega $.  This new energy spectrum is of the form of the quasi-energies discussed in \cite{zeldovich}. If one takes the results of equations \eqref{psi-2} at face value, this would seem to imply a new spectrum with an infinite number of new states labeled by the sideband index $n$. However, from \eqref{psi-2} one finds that the different contributions are weighed by the Bessel functions $J_n (\alpha)$. In Fig. 1, we plot $J_n (\alpha)$ versus $n$ for the example case $\alpha =1000$ . This weighting function oscillates rapidly for $n < \alpha$. At $n \equiv n_{max} \approx |\alpha|$ there is a sharp up shoot, and for $n> n_{max} \approx \alpha$, one finds that $J_n (\alpha)$ exponentially decays to zero so that states beyond $n_{max}$ do not contribute. Due to this weighting factor, $J_n (\alpha)$, the series in \eqref{psi-2} effectively only runs from $n=-n_{max}$ to $n=+n_{max}$. From Fig. 1 one sees that the largest weighting factor, $J_n (\alpha)$, occurs at $n = n_{max} \approx \alpha$.  
\begin{figure}[htb!]
\label{fig1}
    \centering
   \includegraphics[width=0.5\textwidth]{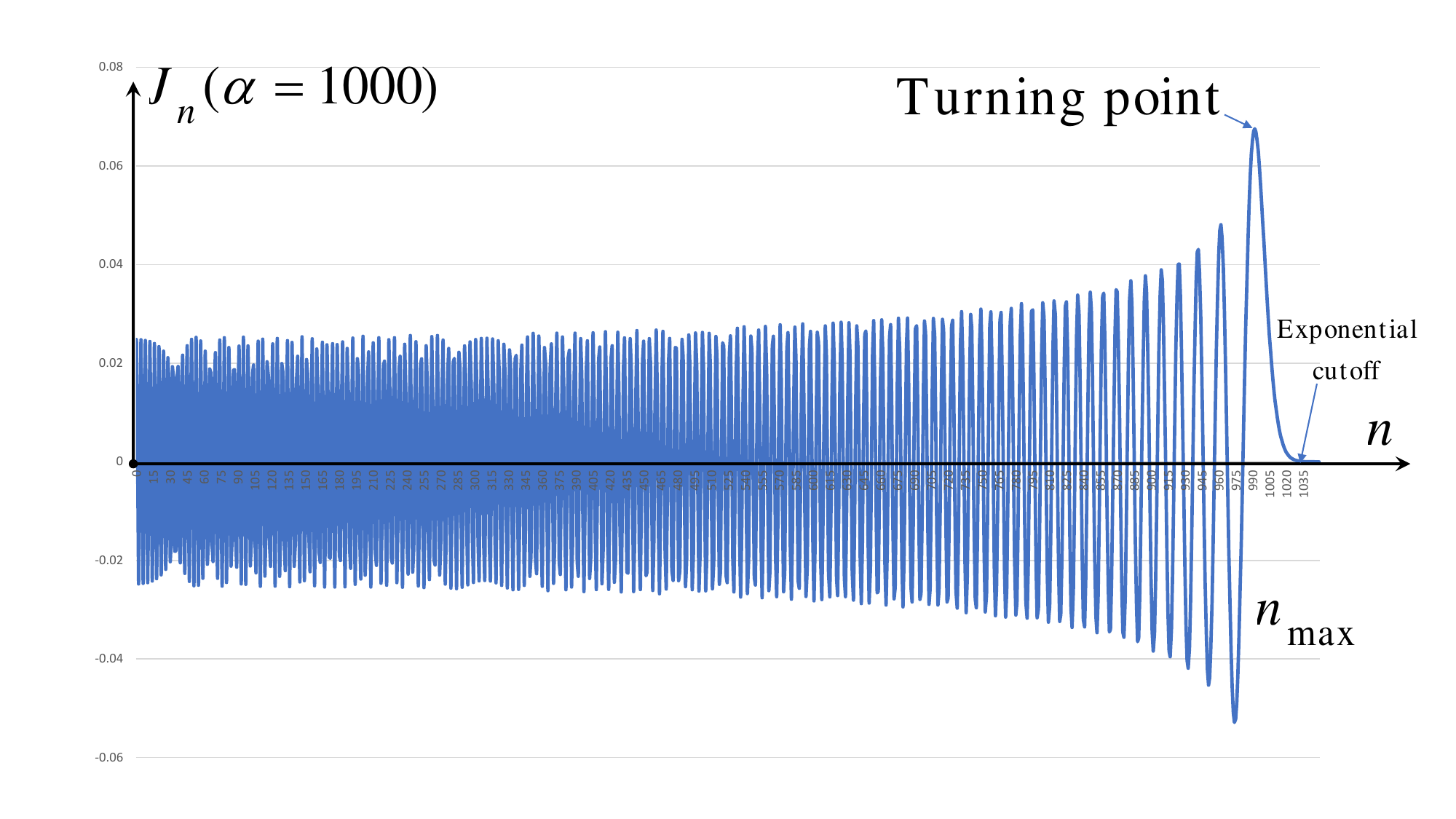}
    \caption{Weighting factor, $J_n (\alpha)$ versus $n$ from \eqref{psi-2} for the example when $\alpha = 1000$. Maximum $J_n (\alpha)$ occurs for $n=n_{max} \approx \alpha$.}
\end{figure}
Combining all this the resulting energy spectrum for the gravitational Hamiltonian, $H$, from \eqref{tdse}  becomes approximately
\begin{equation}
E_i^{(n_{max})} = {\tilde E}_i \pm n_{max} \hbar \Omega ~.
\label{energy-4}
\end{equation}
The original energy $E_i$ has been shifted to ${\tilde E}_i = E_i - \frac{GmM}{A}$ and has developed two dominant energy sidebands $\pm n_{max} \hbar \Omega \approx \pm \alpha \hbar \Omega$. It is these sidebands that are the signal of the gravitational AB effect for this setup. 

In the appendix we show that one can gauge away the sinusodial part of the gravitational potential via a gauge transformation on $\Phi_g(t)$, leaving only the energies  ${\tilde E}_i \equiv E_i - \frac{GmM}{A}$ (the unperturbed energy, $E_i$, plus a constant shift due to gravity). Such a constant shift would not be observable since it would not change the transitions energies {\it i.e.} ${\tilde E}_j - {\tilde E}_i = E_j - E_i$. However the appendix shows that that sinusodial part of the wave-function in \eqref{psi-2} is restored via the gauge transformation of the wave-function $\psi_i$ which comes in conjunction with the gauge transformation of $\Phi_g$. These sidebands are observable since they alter the energy spectrum of the system by adding new levels, rather being an overall shift as is the case for the constant term $- \frac{GmM}{A}$.

There have been other probes of the gravitational AB effect \cite{overstreet}, but these have followed the standard approach of looking for shifts in the interference fringes of quantum particles. Here the signature is a shifting of energy levels, or more precisely the formation of energy sidebands as in \eqref{energy-4}. Next, we discuss how these energy sidebands could be probed for using the M{\"o}ssbauer effect and a temporal version of the Pound-Rebka experiment. 

{\bf M{\"o}ssbauer Effect and Temporal Pound-Rebka Experiment:} We now move to the feasibility of seeing the energy sidebands, $\pm n_{max} \hbar \Omega$, from \eqref{energy-4}. In order for these sidebands to be detectable the width of the lines, $\Delta E$ should be smaller than shift {\it i.e.} $\Delta E < |n_{max} \hbar \Omega|$. As stated at the beginning of the essay we use a temporal version of the Pound-Rebka experiment where the emission and absorption of the $\gamma$-ray photons are temporally separated rather than spatially separated. Using $^{57}$Fe, as in the original experiment \cite{pound}, one has a lifetime of $\tau \sim 10^{-8}$ sec. Thus the emission of the 14.4 keV gamma ray photon has a width of $\Delta E = \frac{\hbar}{\tau} \approx 10^{-8}$ eV. Such small line widths are the strength of using the M{\"o}ssbauer effect. 

The size of the energy sideband shift from \eqref{energy-4} is given by $n_{max} \hbar \Omega$, where $n_{max} \approx \alpha \equiv \frac{GmMB}{\hbar \Omega A^2}$. The energy sidebands are thus $n_{max} \hbar \Omega \approx \frac{GmMB}{A^2}$.  Using the mass of the Earth $M= 5.97 \times 10^{24}$ kg, the nucleon mass $m=1.67 \times 10^{-27}$ kg, the orbital parameters $A=6.805 \times 10^6$ m and $B=-5.0 \times 10^3 $ m, we find that  the size of the energy sidebands are $n_{max} \hbar \Omega \approx \frac{GmMB}{A^2} = 7.18 \times 10^{-23}$ J or $4.49 \times 10^{-4}$ eV. Since this magnitude of the energy sidebands is much greater than the width of $\Delta E \approx 10^{-8}$ eV these sidebands are observable. 

In addition to the sidebands,  $\pm \frac{GmMB}{\hbar \Omega A^2}$, there is also a constant shift of $-\frac{GM}{A}$, to the unperturbed energies, $E_i$. This constant shift is not observable since it does not change the energy differences between levels namely ${\tilde E}_j -{\tilde E}_i = E_j -E_i$. However, the sidebands alter the energy level structure of the system. Thus they do effect the energy level structure of the system which is then observable by scanning through the M{\"o}ssbauer spectrum.  

We note that the shift in energy due to the gravitational interaction, $\frac{GmM}{A}$, is small compared to the original energy, $E_i$. The energy of the original M{\"o}ssbauer $\gamma$-ray is 14.4 keV, while the shift due to the gravitational interaction is $\frac{GmM}{A} \approx 9.8 \times 10^{-20}$ J $\approx 0.6$ eV. We have again used the nucleon mass for $m$ and the Earth's mass for $M$. 

{\bf Summary:} In this essay we proposed a novel approach to testing the gravitational AB effect, which is distinct from the detection of the gravitational AB effect in \cite{overstreet}. The work by Overstreet {\it et al.} used the standard experimental signature for the gravitational AB effect: finding a relative phase shift between two beams of particles ($^{87}$Rb atoms were used in the work of Overstreet {\it et al.}) which were split along different paths and passed through different gravitational potentials. This observation of a phase shift is also the usual way in which the electromagnetic AB effect is observed. Here we propose placing a quantum system in a time-varying gravitational potential and looking for the appearance of energy sidebands. This is the gravitational version of the proposal in \cite{chiao-2023} to probe the scalar-electric AB effect. Due to the weakness of the gravitational interaction, in order to get energy sidebands large enough to observe, we need the time variation of an astrophysical large mass -- in the present essay we use the Earth's mass. This can be achieved by placing the quantum system in a satellite in a low Earth, almost circular orbit. The slight change in the gravitational potential between apogee and perigee provides the change in gravitational potential. Since the satellite is in free fall, this locally eliminates the gravitational field via the equivalence principle. The vanishing of the local gravitational field is a key requirement of tests of the AB effect.     

Starting with \eqref{tdse}, we carried out an analysis parallel to the one used for the scalar-electric AB effect \cite{chiao-2023} but applied it to the gravitational interaction. As in the scalar-electric case, we found that the energy levels of the quantum system developed energy sidebands given in \eqref{energy-4}. In the scalar-electric case, it was relatively easy to change the size of the electric potential and the frequency of the changing electric potential over a wide range. In contrast, for the gravitational case, this was not possible since the variation of the gravitational potential and the frequency are controlled by the parameters of the satellite orbit, which has a narrower range compared to the scalar-electric case. 

In this essay we used a nuclear system via the M{\"o}ssbauer effect for the experimental set-up, to draw a parallel between the gravitational AB effect and the Pound-Rebka test. In \cite{chiao-2024}, it was pointed out that one could also use the Atomic Clock Ensemble in Space (ACES) \cite{aces} system to probe for sidebands, but using an atomic system rather than a nuclear system. 
Just as in the ACES system one would need to run comparisons between the M{\"o}ssbauer system on the satellite and a reference M{\"o}ssbauer system on the ground to observe the energy sidebands. A brief description of how this comparison is carried out is given in reference \cite{aces}. \\

{\bf Acknowledgements}: MET is funded by the ARC Centre of Excellence for Engineered Quantum Systems, Grant No. CE170100009 and the ARC Centre of Excellence for Dark Matter Particle Physics, Grant No. CE200100008
\\

{\bf Appendix }: 

To show that the side bands predicted in this essay are not gauge artifacts we start by gauging the time-dependent Schr{\"o}dinger equation of \eqref{tdse}
\begin{equation}
\label{tdse-1}
i\hbar \frac{\partial \psi }{\partial t}=H\psi =\left( H_{0}+m \Phi_g (t) \right) \psi,
\end{equation}
where $H_0 = -\frac{\hbar ^2}{2m} \nabla ^2 + \Phi _{NG}$ is the Hamiltonian without the gravitational interaction. $\Phi _{NG}$ are the non-gravitational interactions that leads to the eigenfunctions, $\Psi _i ({\bf x})$ and energy eigenvalues, $E_i$, for the time-independent Schr{\"o}dinger equation $H_0 \Psi _i ({\bf x}) = E_i \Psi _i ({\bf x})$ for the system without the gravitational interaction. This is discussed above \eqref{tdse}. We show it is possible to gauge away $\Phi _g (t)$, but one still finds energy side bands. First we expand the metric as $g_{\mu \nu }=\eta _{\mu \nu }+h_{\mu \nu }$, here $\eta _{\mu \nu }$ is flat space-time, Minkowski metric. Following reference \cite{weinberg} we implement a ``gauge transformation" which works on only the perturbations $h_{\mu \nu}$ and have the form 
\begin{equation}
h_{\mu \nu }^{\prime }=h_{\mu \nu } - \partial _{\mu }\xi _{\nu } - \partial
_{\nu }\xi _{\mu }~,  \label{g^mu,nu transformation-1}
\end{equation}
where $\xi ^{\mu }=\left( \xi ^{0},\xi ^{i}\right)$ is an arbitrary displacement four-vector. Equation \eqref{g^mu,nu transformation-1} is the gauge transformation of a spin-2 field, which is a generalization of a spin-1 field gauge transformation $A_\mu ^{\prime }=A_\mu - \partial _{\mu }\lambda$.  Note the spin-2 gauge transformation has four gauge functions, $\xi ^{\mu }=\left( \xi ^{0},\xi ^{i}\right)$, compared to one, $\lambda$, for the spin-1 case. In the Newtonian limit we have $h_{00}\equiv -2\Phi _{g}$, $h_{0i}=0$, and $h_{ij}=0$ (here and throughout the essay we have set $c=1$). Looking at the $00$ component of \eqref{g^mu,nu transformation-1}
\begin{equation}
h_{00 }^{\prime }=h_{00}-\partial _{0}\xi _{0 }-\partial
_{0}\xi _{0} \to \Phi^{\prime} _g = \Phi _g (t) + \partial _t \xi _0~.  \label{g^mu,nu transformation-2}
\end{equation}
We want to shift $h_{00}$, but we want $h_{0i}$, and $h_{ij}$ to remain zero. To accomplish this equation \eqref{g^mu,nu transformation-1} requires we choose $\xi^i=0$ and $\partial _i \xi_0 = 0$. This choice reduces the four gauge functions, $\xi^\mu$, to only one gauge function $\xi^0$ which then makes the gauge transformation of $h_{\mu \nu}$ in this case, very similar to the electromagnetic gauge transformation.  In terms of the analog electromagnetic gauge transformation ({\it i.e.} $\Phi^{\prime} _{EM} = \Phi _{EM} + \partial _t \lambda~$ and ${\bf A}^\prime _{EM} = {\bf A} _{EM}-\nabla \lambda$) this would mean choosing $\nabla \lambda = 0$. Under these conditions the gauge transformation in \eqref{g^mu,nu transformation-1} will leave the off-diagonal and purely spatial components of $h^\prime _{\mu \nu}$ zero {\it i.e.} $h^\prime _{0i}= h^\prime _{ij}=0$. 

To complete the gauge transformation in \eqref{g^mu,nu transformation-2} we also need to add the wave-function transformation namely 
\begin{equation}
\label{gauge-wf}
\psi' _{i }(\mathbf{x},t) = e^{i m \xi_0 /\hbar}\psi _{i }(\mathbf{x},t)~.
\end{equation} 
It is straightforward to check that the combined gauge transformation \eqref{g^mu,nu transformation-2} and \eqref{gauge-wf} leaves the Schr{\"o}dinger equation in \eqref{tdse-1} invariant.

From \eqref{Ab-grav-4} we want to gauge away the time dependent part of $\Phi _g (t)$ namely $ \frac{GmB}{A^2} \cos (\Omega t )$. This is done by choosing $\xi _0 =  - \frac{GMB}{\Omega A^2} \sin (\Omega t ) $ since $\partial _t \xi _0 =  - \frac{GMB}{A^2} \cos (\Omega t )$. With these choices we have $\Phi ^\prime _g = \Phi _g + \partial _t \xi ^0 = -\frac{GM}{A}$ in \eqref{g^mu,nu transformation-2} {\it i.e.} the gravitational potential in this gauge is a constant. 

Since we have gauged away the time varying part of gravitational potential it seems this would make the effect go away. However, the gauge transformation in \eqref{g^mu,nu transformation-2} also requires that the wave-function $\psi_i ({\bf x}, t) =\Psi _{i }(\mathbf{x})\exp \left( -\frac{i}{\hbar} (E_i -GM/A) t\right)$) also be transformed combination with the transformation is \eqref{g^mu,nu transformation-2}. Note we have included the constant gravitational shift, $-\frac{Gm}{A}$, with the non-gravitational energy, $E_i$. This constant shift of all the energies $E_i$ means that this shift of $-\frac{Gm}{A}$ is unobservable since it shifts all the energies by the same amount.  Applying the wave-function gauge transformation in \eqref{gauge-wf} gives
\begin{eqnarray}
\psi' _{i }(\mathbf{x},t) &=& \exp \left(  \frac{i m}{\hbar} \xi_0 \right) \Psi _{i }(\mathbf{x})\exp \left( -\frac{i}{\hbar}
\left({ E}_i - \frac{GmM}{A} \right) t\right) \nonumber \\
&=&  \exp \left( - i \frac{GmMB}{\hbar \Omega A^2} \sin (\Omega t )  \right) \Psi _{i }(\mathbf{x})\exp \left( -\frac{i}{\hbar}
\left({ E}_i - \frac{GmM}{A} \right) t \right)
~. 
\label{psi-1}
\end{eqnarray}
Comparing the wave-function in \eqref{psi-1}, where the gravitational potential has been gauged away, with \eqref{T(t) solution} and \eqref{psi} without the gravitational interaction gauged away, one finds they are the same. In the case were the gravitational potential has been gauged away, one nevertheless obtains the same result due to the transformation of the wave-function in the first line of \eqref{psi-1}. 
This argument is parallel to that for the scalar electric case at the of section II in \cite{chiao-2023}.

The results of \eqref{psi-1}, or equivalently \eqref{T(t) solution} and \eqref{psi}, show that the gravitational potential has two effects on the wave-function and energies, $E_i$. First there is a constant shift, $-\frac{GmM}{A}$. This term is not observable since it shifts all of the $E_i$'s by the same amount and thus does not lead to any shift in transitions between energy levels {\it i.e.} $E_j - E_i$ is not altered from the constant term $-\frac{GmM}{A}$. The time dependent, sinusodial term gives rise to energy side bands, which are obserable since they change the energy level structure of the system.

\end{document}